\documentclass[aps,pra,twocolumn,superscriptaddress,showpacs,floatfix]{revtex4}
\usepackage{amsmath}
\usepackage{amssymb}
\usepackage{graphicx,color,hyperref}
\usepackage{latexsym}
\usepackage{graphics}
\usepackage{subfigure}

\def\be{\begin{equation}}
\def\ee{\end{equation}}
\def\bea{\begin{eqnarray}}
\def\eea{\end{eqnarray}}
\def\bse{\begin{subequations}}
\def\ese{\end{subequations}}

\def\be{\begin{eqnarray}}
\def\ee{\end{eqnarray}}

\begin{document}

\title{Creating a giant and tunable spin squeezing via a time-dependent
collective atom-photon coupling}
\author{Lixian Yu}
\affiliation{Department of Physics, Shaoxing University, Shaoxing 312000, China}
\affiliation{State Key Laboratory of Quantum Optics and Quantum Optics Devices, Institute
of Laser spectroscopy, Shanxi University, Taiyuan 030006, China}
\author{Jingtao Fan}
\affiliation{State Key Laboratory of Quantum Optics and Quantum Optics Devices, Institute
of Laser spectroscopy, Shanxi University, Taiyuan 030006, China}
\author{Shiqun Zhu}
\affiliation{School of Physical Science and Technology, Soochow University, Suzhou
215006, China}
\author{Gang Chen}
\thanks{chengang971@163.com}
\affiliation{State Key Laboratory of Quantum Optics and Quantum Optics Devices, Institute
of Laser spectroscopy, Shanxi University, Taiyuan 030006, China}
\author{Suotang Jia}
\thanks{tjia@sxu.edu.cn}
\affiliation{State Key Laboratory of Quantum Optics and Quantum Optics Devices, Institute
of Laser spectroscopy, Shanxi University, Taiyuan 030006, China}
\author{Franco Nori}
\thanks{fnori@riken.jp}
\affiliation{CEMS, RIKEN, Saitama 351-0198, Japan}
\affiliation{Physics Department, The University of Michigan, Ann Arbor, Michigan
48109-1040, USA}
\affiliation{Department of Physics, Korea University, Seoul 136-713, Korea}

\begin{abstract}
We present an experimentally-feasible method to produce a giant and tunable
spin squeezing, when an ensemble of many four-level atoms interacts
simultaneously with a single-mode photon and classical driving lasers. Our
approach is to simply introduce a time-dependent collective atom-photon
coupling. We show that the maximal squeezing factor measured experimentally
can be well controlled by both its driving magnitude and driving frequency.
Especially, when increasing the driving magnitude, the maximal squeezing
factor increases, and thus can be enhanced rapidly. We also demonstrate
explicitly, in the high-frequency approximation, that this spin squeezing
arises from a strong repulsive spin-spin interaction induced by the
time-dependent collective atom-photon coupling. Finally, we evaluate
analytically, using current experimental parameters, the maximal squeezing
factor, which can reach $40$ dB. This giant squeezing factor is far larger
than previous ones.
\end{abstract}

\maketitle

Spin squeezing states are quantum correlated states with reduced
fluctuations in one of the collective spin components \cite{MK93,JM11,NPR13}%
. Such states not only play a central role in investigating many-body
entanglement \cite{JM11,Dl00,AS01,JKK05,GT09,OG09,RJS12}, but also have
possible applications in atom interferometers and high-precision atom clocks
\cite{JM11,SG99,KH10}. Now the preparation of spin squeezing states has
become an important subject in quantum information and quantum metrology
\cite{JM11,NPR13}. In principle, nonlinear spin-spin interactions are
necessary for producing spin squeezing states, and moreover, have been
constructed experimentally in both multicomponent Bose-Einstein condensates
(BECs) \cite{JE08,GR10,MFR10,EMB11,BL11,CDH12,MJM13} and atom-cavity
interacting systems \cite{IDL10,ZS11}. However, the generated spin-spin
interactions are weak, and thus the corresponding maximal squeezing factors
(MSFs) acquired are lower than $10$ dB \cite{JM11,NPR13}. Recently, many
proposals \cite{XH97,XH99,AMZ08,COL10,TV11,YCL11,BJ12,AZC12,CS13,JL13} have
been suggested to enhance the upper limits of the MSFs in laboratory
conditions, but the experimental challenges are difficult.

Here we present an experimentally-feasible method to achieve a giant and
tunable spin squeezing, when an ensemble of many four-level atoms interacts
simultaneously with a single-mode photon and classical driving lasers.
Recently, a similar setup has been considered experimentally in a BEC-cavity
system, and a remarkable quantum phase transition, from a normal phase to a
superradiant phase of the Dicke model, was observed \cite{KB10,KB11}. The
distinct advantage of this setup is that the realized Dicke model has a
tunable collective atom-photon coupling through manipulating the intensities
of the classical driving lasers \cite{FD07}.

The central idea of our work is to simply introduce a time-dependent
collective atom-photon coupling in the realized Dicke model. We show that
the MSF can be well controlled by both its driving magnitude and driving
frequency. In particular, when increasing the driving magnitude, the MSF
increases, in contrast to the known results of the undriven Dicke model \cite%
{JM11}, and thus can be enhanced rapidly. In the high-frequency
approximation, we demonstrate explicitly that this spin squeezing arises
from a strong repulsive spin-spin interaction induced by the time-dependent
collective atom-photon coupling (for the undriven Dicke model, only a weak
attractive spin-spin interaction is generated). Finally, we evaluate
analytically, using current experimental parameters \cite{KB10,KB11}, the
MSF, which can reach $40$ dB. This giant MSF is far larger than previous
ones \cite{JE08,GR10,MFR10,EMB11,BL11,CDH12,MJM13,IDL10,ZS11}.

\begin{figure}[tp]
\includegraphics[width=8.5cm]{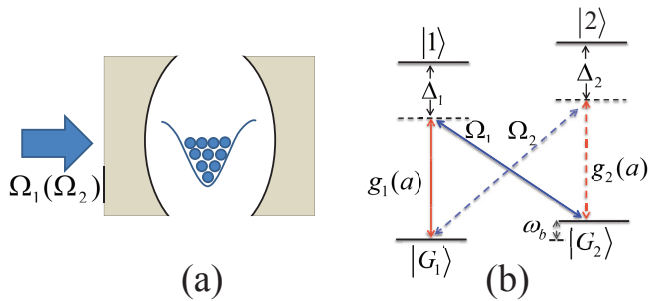}\newline
\caption{(Color online)(a) Proposed experimental setup. (b) The atom energy
levels, which are driven simultaneously by a single-mode photon of the
optical cavity and a pair of classical driving lasers.}
\label{fig1}
\end{figure}

\section{Model and Hamiltonian}

Figure \ref{fig1}a shows our proposed experimental setup, in which an
ensemble of many four-level atoms interacts simultaneously with a
single-mode photon of the optical cavity and a pair of classical driving
lasers. Each atom has two stable ground states, labeled respectively by $%
\left\vert G_{1}\right\rangle $ and $\left\vert G_{2}\right\rangle $, which
are coupled through a pair of Raman channels, as shown in Fig. \ref{fig1}b.
The photon, with the creation and annihilation operators $a^{\dagger }$ and $%
a$, mediates the $\left\vert G_{1}\right\rangle \longleftrightarrow
\left\vert 1\right\rangle $ and $\left\vert G_{2}\right\rangle
\longleftrightarrow \left\vert 2\right\rangle $ transitions, with
atom-photon coupling strengths $g_{1}$ and $g_{2}$, whereas the classical
driving lasers induce the $\left\vert G_{1}\right\rangle \longleftrightarrow
\left\vert 2\right\rangle $ and $\left\vert G_{2}\right\rangle
\longleftrightarrow \left\vert 1\right\rangle $ transitions, with Rabi
frequencies $\Omega _{1}\ $and $\Omega _{2}$.

In the large-detuning limit, the excited states of the atoms can be
eliminated adiabatically, and thus an effective two-level system, with the
collective spin operators $S_{x}=\sum_{i}(\left\vert G_{2}\right\rangle
_{ii}\left\langle G_{1}\right\vert +\left\vert G_{1}\right\rangle
_{ii}\left\langle G_{2}\right\vert )$ and $S_{z}=\sum_{i}(\left\vert
G_{2}\right\rangle _{ii}\left\langle G_{2}\right\vert -\left\vert
G_{1}\right\rangle _{ii}\left\langle G_{1}\right\vert )$, can be
constructed. When the parameters are chosen as \cite{FD07}
\begin{equation}
\frac{g_{1}^{2}}{\Delta _{1}}=\frac{g_{2}^{2}}{\Delta _{2}}\text{, }\frac{%
g_{1}\Omega _{1}}{\Delta _{1}}=\frac{g_{2}\Omega _{2}}{\Delta _{2}},
\label{PA}
\end{equation}%
we realize a Dicke-like Hamiltonian \cite{DH54}
\begin{equation}
H=\Delta _{p}a^{\dagger }a+\omega _{0}S_{z}+\frac{g}{\sqrt{N}}(a^{\dagger
}+a)S_{x}.  \label{Total H}
\end{equation}%
In the Hamiltonian (\ref{Total H}), the effective cavity frequency $\Delta
_{p}=\delta _{c}+Ng_{1}^{2}/\Delta _{1}$, where $\delta _{c}=\omega
_{c}-(\omega _{l2}-\omega _{b}^{\prime })$, $N$ is the number of atoms, $%
\omega _{c}$ is the real cavity frequency, $\omega _{b}^{\prime }=(\omega
_{l2}-\omega _{l1})/2$ is a frequency close to the frequency $\omega _{b}$
of the energy level $\left\vert G_{2}\right\rangle $, $\omega _{l1}$ and $%
\omega _{l2}$ are the driving frequencies of the classical lasers,
respectively, and $\Delta _{1}$ is the detuning. This effective cavity
frequency $\Delta _{p}$ varies from -GHz to GHz, and even goes beyond this
regime. However, when $\Delta _{p}<0$, the system becomes unstable \cite%
{KB10,KB11}. Thus, hereafter we consider $\Delta _{p}\geq 0$. The effective
atom frequency $\omega _{0}=(\omega _{b}-\omega _{b}^{\prime })$, which is
of the order of several hundred kHz. The collective atom-photon coupling
strength $g=\sqrt{N}g_{1}\Omega _{1}/\Delta _{1}$, which can reach the order
of a GHz by independently manipulating the intensities of the classical
driving lasers.

\begin{figure}[tp]
\includegraphics[width=8.5cm]{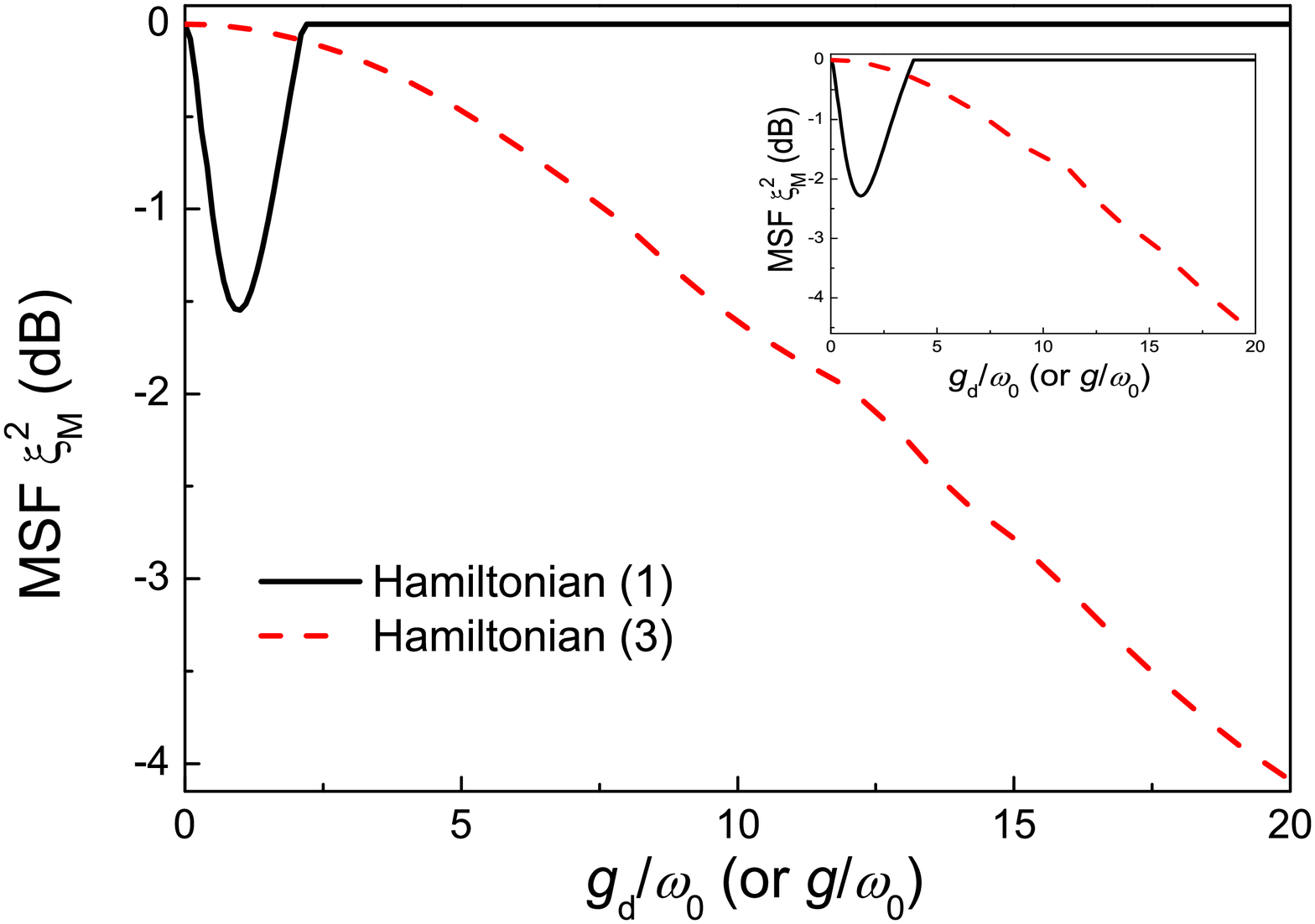}\newline
\caption{(Color online) Numerical plot of the MSF $\protect\xi _{\text{M}%
}^{2}$ of the undriven Dicke model (\protect\ref{Total H}) (black solid
curve) and the time-dependent Hamiltonian (\protect\ref{HT}) (red dashed
curve), when $N=10$. Inset: The MSF $\protect\xi _{\text{M}}^{2}$ for $N=20$%
. In these figures, the other parameters are chosen as $\Delta _{p}=\protect%
\omega _{0}$ and $\protect\omega =10\protect\omega _{0}$.}
\label{fig2}
\end{figure}

\section{Spin squeezing}

This four-level model has been regarded as a promising candidate to produce
both field squeezing and spin squeezing. For example, two-mode field
squeezing \cite{RG06} and unconditional two-mode squeezing of separated
atomic ensembles \cite{ASP06} have been considered by introducing two
cavities, mediating the $\left\vert G_{1}\right\rangle \longleftrightarrow
\left\vert 1\right\rangle $ and $\left\vert G_{2}\right\rangle
\longleftrightarrow \left\vert 2\right\rangle $ transitions, respectively.
Recently, it has been proposed that spin squeezing can be achieved by
designing degenerate ground states $\left\vert G_{1}\right\rangle $ and $%
\left\vert G_{2}\right\rangle $ ($\omega _{0}=0$) \cite{SB12,EG13}.
Especially, Ref. \cite{EG13} demonstrated the existence of a collective
atomic dark state, decoupled from the cavity mode field. When explicitly
constructing this steady dark state, spin squeezing, which is considerably
more robust against noise, can be achieved simultaneously \cite{EG13}. Here,
we mainly achieve a giant and tunable spin squeezing by considering a
time-dependent collective atom-photon coupling strength $g(t)$, and explore
its physical consequences.

When the Rabi frequencies of the classical driving lasers are chosen as
\begin{equation}
\Omega _{1}=\Omega _{2}=\Omega _{d}\cos (\omega t),  \label{TRAB}
\end{equation}%
the collective atom-photon coupling strength becomes
\begin{equation}
g(t)=g_{_{d}}\cos (\omega t)\text{,}  \label{TDCO}
\end{equation}%
where $g_{_{d}}=\sqrt{N}g_{1}\Omega _{d}/\Delta _{1}$ is the effective
driving magnitude and $\omega $ is the driving frequency. Substituting Eq. (%
\ref{TDCO}) into the Hamiltonian (\ref{Total H}) yields a time-dependent
Dicke model
\begin{equation}
H(t)=\Delta _{p}a^{\dagger }a+\omega _{0}S_{z}+\frac{g_{_{d}}\cos (\omega t)%
}{\sqrt{N}}(a^{\dagger }+a)S_{x}.  \label{HT}
\end{equation}%
If the initial state is chosen as
\begin{equation}
\left\vert \psi (0)\right\rangle =\left\vert S_{z}=-\frac{N}{2}\right\rangle
\otimes \left\vert 0\right\rangle ,  \label{INS}
\end{equation}%
the corresponding squeezing factor is defined as \cite{DJW94}
\begin{equation}
\xi _{\text{R}}^{2}(t)=\frac{N\Delta S_{\vec{n}_{\bot }}^{2}(t)}{\left\vert
S(t)\right\vert ^{2}},  \label{SF}
\end{equation}%
where $\vec{n}_{\bot }$ refers to an axis, which is perpendicular to the
mean-spin direction, $\left\vert S\right\vert =\sqrt{\left\langle
S_{x}\right\rangle ^{2}+\left\langle S_{y}\right\rangle ^{2}+\left\langle
S_{z}\right\rangle ^{2}}$, and
\begin{equation}
\Delta A^{2}=\left\langle A^{2}\right\rangle -\left\langle A\right\rangle
^{2}  \label{DETA}
\end{equation}%
is the standard deviation. If $\left\vert \xi _{\text{R}}^{2}\right\vert <1$%
, the state is spin squeezed, and its phase sensitivity, $\Delta \varphi
=\xi _{\text{R}}^{2}/\sqrt{N}$, is improved over the shot-noise limit. In
addition, the initial state $\left\vert 0\right\rangle $ is a pure state not
containing any photons and consequently is not affected by the cavity decay
\cite{XW10,XY12}.

It is very difficult to derive an analytical expression for the squeezing
factor of the time-dependent Hamiltonian (\ref{HT}). However, in
experiments, the MSF
\begin{equation}
\xi _{\text{M}}^{2}=\min [\xi _{\text{R}}^{2}(t)]  \label{SSQM}
\end{equation}%
is usually measured \cite{JM11}. Thus, hereafter we focus mainly on this MSF
$\xi _{\text{M}}^{2}$. In Fig. \ref{fig2}, we numerically calculate the MSF $%
\xi _{\text{M}}^{2}$ of the undriven Dicke model (\ref{Total H}) (black
solid curve) and the time-dependent Hamiltonian (\ref{HT}) (red dashed
curve), with the same initial state $\left\vert \psi (0)\right\rangle $.

For the undriven Dicke model (\ref{Total H}), when increasing the static
collective atom-photon coupling $g$, the MSF $\xi _{\text{M}}^{2}$ increases
rapidly, and then decreases once $g$ goes beyond a critical value $g_{c}$.
Its physics can be understood as follows. When $g<g_{c}$, the system is
located at the normal phase with no macroscopic collective excitations of
both the atoms and the photon, i.e., $\left\langle a^{\dagger
}a\right\rangle =0$. However, this virtual photon acts as a bus, and thus
generates an attractive spin-spin interaction $-S_{x}^{2}$, which can be
demonstrated, in the limit when $\Delta _{p}\gg g$, by a second-order
perturbation theory \cite{SM08,GC09,JL10,SDB13}. Moreover, the interaction
strength depends on $g^{2}/\Delta _{p}$. Thus, when increasing a weak $g$,
the MSF $\xi _{\text{M}}^{2}$ increases. When $g>g_{c}$, the undriven Dicke
model exhibits a strong atom-photon interaction. When increasing $g$, the
atoms and the photon become more and more entangled, and the spin squeezing
is suppressed. Especially, in the limit when $g\gg \{\Delta _{p},\omega
_{0}\}$, this atom-photon interaction plays a dominate role in the quantum
dynamics of the undriven Dicke model. For the given initial state $%
\left\vert \psi (0)\right\rangle $, we have
\begin{equation}
S_{z}(t)=\frac{N}{2}\cos (Cgt),  \label{SZT}
\end{equation}%
where $C=a^{\dag }+a$, and thus
\begin{equation}
\xi _{\text{R}}^{2}(t)=0.  \label{MS}
\end{equation}%
This result agrees well with the direct numerical calculation, as shown by
the black solid curve of Fig. \ref{fig2}.

\begin{figure}[tp]
\includegraphics[width=8.5cm]{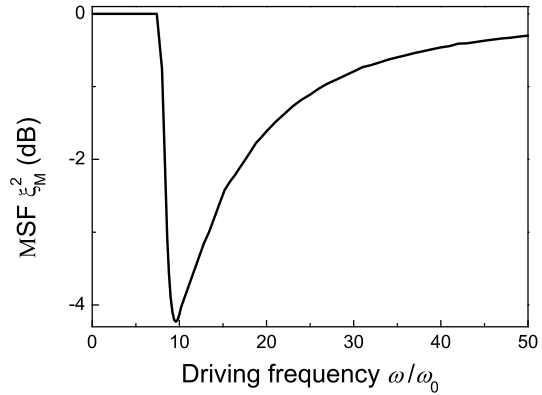}\newline
\caption{(Color online) Numerical plot of the MSF $\protect\xi _{\text{M}%
}^{2}$ of the time-dependent Hamiltonian (\protect\ref{HT}), when $g_{d}=20%
\protect\omega _{0}$ and $N=10$.}
\label{fig3}
\end{figure}

For the time-dependent Hamiltonian (\ref{HT}), the MSF $\xi _{\text{M}}^{2}$
exhibits some surprising behaviors. As shown by the red dashed curve of Fig. %
\ref{fig2}, the MSF $\xi _{\text{M}}^{2}$ can be largely enhanced by
increasing the driving magnitude $g_{_{d}}$, which is in contrast to the
results of the undriven Dicke model. In Fig. \ref{fig3}, we numerically plot
the MSF $\xi _{\text{M}}^{2}$ as a function of the driving frequency $\omega
$. We find that the MSF $\xi _{\text{M}}^{2}$ can also be enhanced by
choosing a proper driving frequency $\omega $. In the high-frequency regime,
the MSF $\xi _{\text{M}}^{2}$ decreases when increasing the driving
frequency $\omega $. The above predictions imply that a giant MSF can be
prepared by controlling the time-dependent collective atom-photon coupling $%
g(t)$ in experiments.

\section{$g_{_{d}}-$induced strong repulsive spin-spin interaction}

We now illustrate the fundamental physics why these surprising behaviors of
spin squeezing can occur in the driven Dicke model. In general, we cannot
extract the interesting physics for any driving frequency $\omega $.
Fortunately, in the high-frequency approximation, we will demonstrate
explicitly that the time-dependent collective atom-photon coupling gives
rise to a magnitude-dependent repulsive spin-spin interaction, which is
essential for producing spin squeezing. This result is quite different from
that of the undriven Dicke model, in which only a weak attractive spin-spin
interaction is generated by the static collective coupling.

We first employ a time-dependent unitary transformation
\begin{equation}
U\left( t\right) =\exp [-i\chi \sin (\omega t)\left( a^{\dag }+a\right)
S_{x}],  \label{UO}
\end{equation}%
with
\begin{equation}
\chi =\frac{g_{d}}{\omega \sqrt{N}},  \label{PXS}
\end{equation}%
to rewrite the time-dependent Hamiltonian (\ref{HT}) as
\begin{equation}
H_{\text{u}}(t)=U^{\dag }(t)H(t)U(t)-iU^{\dag }(t)\frac{\partial U(t)}{%
\partial t}.  \label{HUT}
\end{equation}%
After a straightforward calculation, we have
\begin{widetext}
\begin{equation}
H_{\text{u}}(t)=\Delta _{p}[a^{\dag }a+i\chi \sin (\omega t)(-a^{\dag }+a)S_{x}+\chi
^{2}\sin ^{2}(\omega t)S_{x}^{2}]+\omega _{0}\{S_{z}\cos [\chi \sin (\omega
t)(a^{\dag }+a)]+S_{y}\sin [\chi \sin (\omega t)(a^{\dag }+a)]\}.
\label{HTND}
\end{equation}
\end{widetext}In addition, for a given quantum state $\left\vert \psi
(t)\right\rangle $\ of the Hamiltonian (\ref{HT}), the time-dependent
quantum state of the Hamiltonian (\ref{HUT}) is written as
\begin{equation}
\left\vert \psi _{\text{u}}(t)\right\rangle =U(t)\left\vert \psi
(t)\right\rangle .  \label{STA}
\end{equation}%
When $t=0$, $\left\vert \psi _{\text{u}}(0)\right\rangle =U(0)\left\vert
\psi (0)\right\rangle =\left\vert \psi (0)\right\rangle $.

By means of the formulas
\begin{equation}
\left\{
\begin{array}{c}
\cos [\vartheta \sin (\omega t)]=J_{0}(\vartheta )+2\sum_{m=1}^{\infty
}J_{2m}(\vartheta )\cos (2m\omega t) \\
\sin [\vartheta \sin (\omega t)]=2\sum_{m=1}^{\infty }J_{2m+1}(\vartheta
)\sin [(2m+1)\omega t]%
\end{array}%
\right. ,  \label{FOR}
\end{equation}%
where $J_{0}(\cdot )$ and $J_{m}(\cdot )$ are the zeroth- and integer- order
Bessel functions, respectively, the time-dependent Hamiltonian (\ref{HTND})
is rewritten as
\begin{equation}
H_{\text{u}}(t)=\sum\limits_{n=-\infty }^{\infty }h_{n}\exp (in\omega t),
\label{EHT}
\end{equation}%
where
\begin{equation}
h_{-1}=\omega _{0}J_{1}\left[ \frac{g_{d}(a^{\dag }+a)}{\sqrt{N}\omega }%
\right] S_{y}-\frac{\Delta _{p}g_{d}(a-a^{\dag })S_{x}}{2\sqrt{N}\omega },
\label{h1}
\end{equation}%
\begin{equation}
h_{0}=\Delta _{p}a^{\dag }a+\omega _{0}J_{0}\left[ \frac{g_{d}(a^{\dag }+a)}{%
\sqrt{N}\omega }\right] S_{z}+\frac{\Delta _{p}g_{d}^{2}S_{x}^{2}}{2N\omega
^{2}},  \label{h2}
\end{equation}%
\begin{equation}
h_{1}=\frac{\Delta _{p}g_{d}(a-a^{\dag })S_{x}}{2\sqrt{N}\omega }+\omega
_{0}J_{1}\left[ \frac{g_{d}(a^{\dag }+a)}{\sqrt{N}\omega }\right] S_{y}.
\label{h3}
\end{equation}%
The other expressions for $h_{n}$ ($n\geq 2$) are too complicated to list
here.

\begin{figure}[tp]
\includegraphics[width=8.5cm]{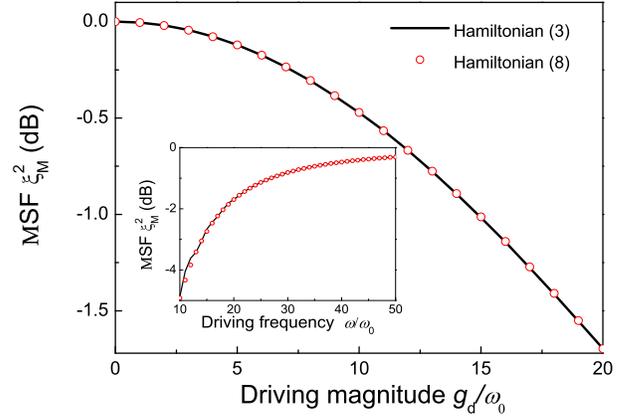}\newline
\caption{Numerical plot of the MSF $\protect\xi _{\text{M}}^{2} $ as a
function of the driving magnitude $g_{d}$, when $\protect\omega=20\protect%
\omega _{0}$ and $N=10$. Insert: The MSF $\protect\xi _{\text{M}}^{2}$
versus the driving frequency $\protect\omega$, when $g_{d}=20\protect\omega %
_{0}$ and $N=10$. In these figures, the black solid curves denote the
results of the time-dependent Hamiltonian (\protect\ref{HT}) and the red
open symbols correspond to the results of the effective time-independent
Hamiltonian (\protect\ref{EHAM}). }
\label{fig4}
\end{figure}

In the high-frequency approximation ($\omega \gg \{\Delta _{p},\omega _{0}\}$%
) \cite{MG98}, we neglect all the time-dependent terms in the Hamiltonian (%
\ref{EHT}), in analogy with the standard rotating-wave approximation, and
then obtain an effective time-independent Hamiltonian
\begin{equation}
H_{\text{e}}=\frac{q}{N}S_{x}^{2}+\Delta _{p}a^{\dag }a+\omega _{0}J_{0}%
\left[ \frac{g_{d}}{\sqrt{N}\omega }(a^{\dag }+a)\right] S_{z},  \label{EHAM}
\end{equation}%
where
\begin{equation}
q=\frac{\Delta _{p}g_{d}^{2}}{2\omega ^{2}}.  \label{PAQ}
\end{equation}%
The Hamiltonian (\ref{EHAM}) shows clearly that the time-dependent
collective atom-photon coupling induces a repulsive spin-spin interaction ($%
q>0$ for $\Delta _{p}>0$), which can be controlled widely and independently
by tuning the effective cavity frequency $\Delta _{p}$, and especially, the
driving magnitude $g_{d}$ and the driving frequency $\omega $.

We emphasize that for the undriven Dicke model, the attractive spin-spin
interaction $-S_{x}^{2}$\ is mediated by a virtual photon. As a result, its
interaction strength is weak, and can be derived from second-order
perturbation theory when $\Delta _{p}\gg g$\ \cite{SM08,GC09,JL10,SDB13}.
However, the repulsive spin-spin interaction realized here arises from the
driving photon under the high-frequency approximation, which needs to
satisfy the following condition: $\omega \gg \{\Delta _{p},\omega _{0}\}$\
(the condition $\Delta _{p}\gg g$\ in the undriven Dicke model is now
relaxed). This means that the driving magnitude $g_{_{d}}$\ can reach the
same order as the driving frequency $\omega $,\ and go beyond the effective
cavity frequency $\Delta _{p}$. Thus, the corresponding interaction strength
can reach a large value. For example, when the parameters are chosen as $%
g_{d}=\omega =2\pi \times 0.5 $\ GHz and $\Delta _{p}=0.1\omega =2\pi \times
0.05$\ GHz, then the repulsive spin-spin interaction strength becomes $%
q=\Delta _{p}g_{d}^{2}/(2\omega ^{2})=2\pi \times 250$\ MHz, which is 2-3
orders larger than that of the undriven Dicke model \cite{IDL10,ZS11}.

In order to further reveal the role of the generated repulsive spin-spin
interaction $q$, in Fig. \ref{fig4}, we numerically compare the MSF $\xi _{%
\text{M}}^{2}$ of the time-dependent Hamiltonian (\ref{HT}) with that of the
effective time-independent Hamiltonian (\ref{EHAM}). These results imply
that the spin squeezing of the time-dependent Hamiltonian (\ref{HT}) for the
initial state $\left\vert \psi (0)\right\rangle $ can be well described by
the effective Hamiltonian (\ref{EHAM}) in the high-frequency approximation.
That is, we can employ the effective time-independent Hamiltonian (\ref{EHAM}%
) to analyze the predictions in Figs. \ref{fig2} and \ref{fig3}.

It should be remarked that for the time-independent Hamiltonian (\ref{EHAM}%
), there also exists a weak photon-induced spin-spin interaction in the $z$
direction, apart from the repulsive spin-spin interaction $qS_{x}^{2}/N$.
However, when the initial state is chosen as $\left\vert \psi
(0)\right\rangle =\left\vert S_{z}=-N/2\right\rangle \otimes \left\vert
0\right\rangle $, this photon-induced spin-spin interaction has almost no
role in producing spin squeezing. This means that the repulsive spin-spin
interaction is central for producing spin squeezing in the time-dependent
Hamiltonian (\ref{HT}), with the initial state $\left\vert \psi
(0)\right\rangle $. When increasing the driving magnitude $g_{_{d}}$, this
repulsive spin-spin interaction $q$ increases, and reaches a large value.
This strong repulsive spin-spin interaction $q$ can significantly enhance
the MSF $\xi _{\text{M}}^{2}$, as shown by the red dashed curve of Fig. \ref%
{fig2}. However, when increasing the driving frequency $\omega $, the
repulsive spin-spin interaction $q$ becomes weaker, and correspondingly, the
MSF $\xi _{\text{M}}^{2}$ decreases, as shown in Fig. \ref{fig3}.

\begin{figure}[tp]
\includegraphics[width=8cm]{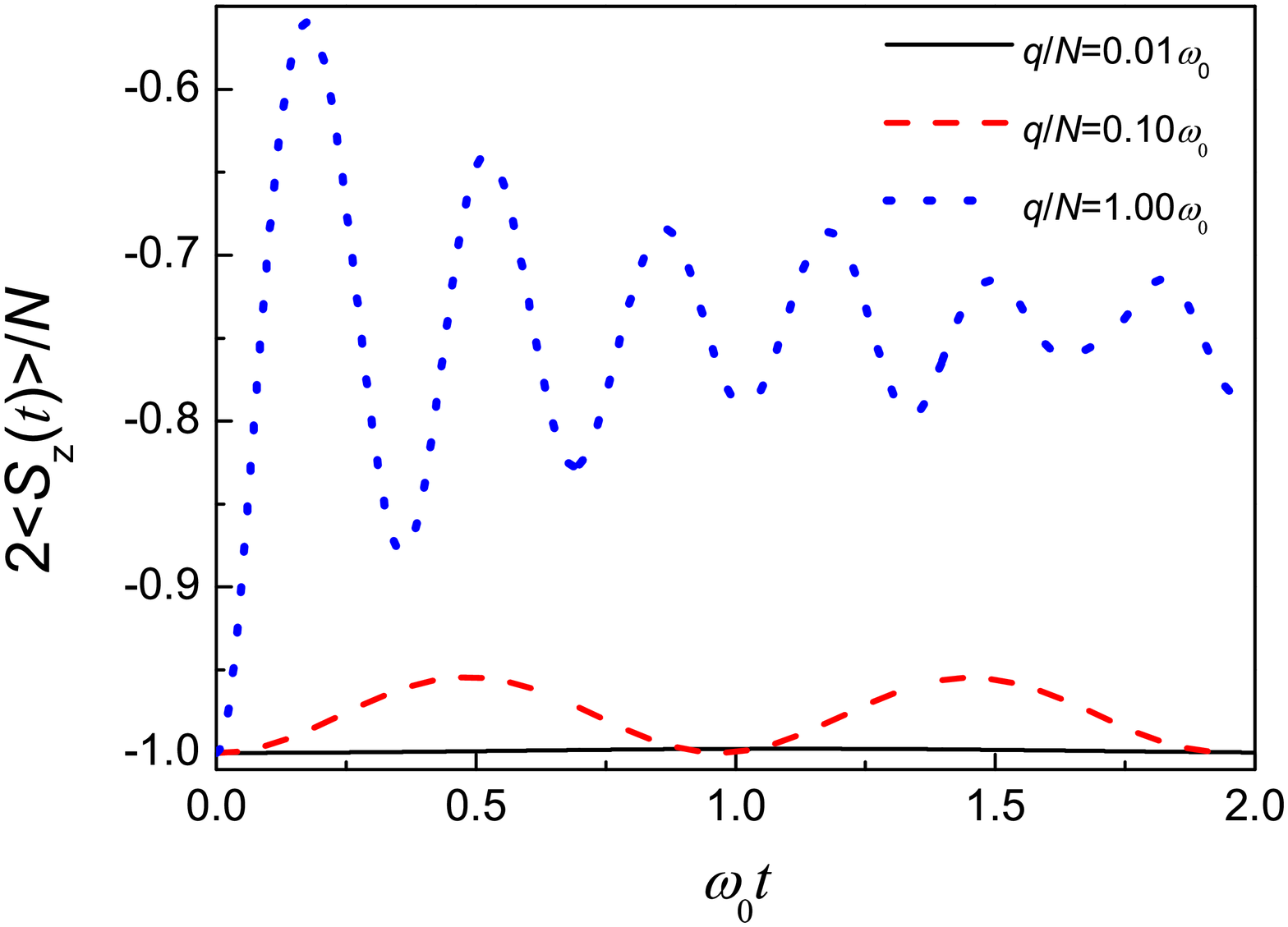}\newline
\caption{(Color online) Quantum dynamics of $2\left\langle
S_{z}(t)\right\rangle /N$ for the different spin-spin interaction strengths:
$q/N=0.01\protect\omega _{0},q/N=0.1\protect\omega _{0}$, and $q/N=\protect%
\omega _{0}$, when $N=100$.}
\label{fig5}
\end{figure}

\section{A giant squeezing factor in experiments}

In this section, we evaluate the MSF $\xi _{\text{M}}^{2}$ by considering
current experimental parameters, especially with a large atom number. For a
large atom number $N\sim 10^{4}$, the MSF $\xi _{\text{M}}^{2}$ is hard to
obtain numerically. Fortunately, in such a case, we have $g_{d}/(\sqrt{N}%
\omega )\rightarrow 0$. This implies that the effective time-independent
Hamiltonian (\ref{EHAM}) becomes
\begin{equation}
H_{\text{e}}=\frac{q}{N}S_{x}^{2}+\omega _{0}S_{z}.  \label{HE}
\end{equation}%
When $\omega _{0}\gg q/N$, the MSF $\xi _{\text{M}}^{2}$ for the Hamiltonian
(\ref{HE}) can be derived explicitly from the frozen-spin approximation \cite%
{JM11}.

In terms of the Heisenberg equation of motion, we obtain
\begin{equation}
\left\{
\begin{array}{c}
\dot{S}_{x}=-\omega _{0}S_{y}, \\
\dot{S}_{y}=-\frac{q}{N}(S_{z}S_{x}+S_{x}S_{z})+\omega _{0}S_{x}.%
\end{array}%
\right.  \label{S1}
\end{equation}%
For the given initial state $\left\vert \psi (0)\right\rangle =\left\vert
S_{z}=-\frac{N}{2}\right\rangle $, $\left\langle S_{y}(0)\right\rangle
=\left\langle S_{x}(0)\right\rangle =0$ and $\left\langle
S_{y}^{2}(0)\right\rangle =\left\langle S_{x}^{2}(0)\right\rangle =N/4$. In
general, the differential equations (\ref{S1})) cannot be solved
analytically. However, when $\omega _{0}\gg q/N$, $2\left\langle
S_{z}(t)\right\rangle /N$ remains approximately unchanged under the initial
state $\left\vert \psi (0)\right\rangle $, as shown in Fig. \ref{fig5}. This
implies that we can make an approximation by replacing $S_{z}$ by $-N/2$,
which leads to the harmonic solutions
\begin{equation}
\left\{
\begin{array}{c}
S_{x}(t)\simeq S_{x}(0)\cos (\eta t)+\frac{\omega _{0}}{\eta }S_{y}(0)\sin
(\eta t), \\
S_{y}(t)\simeq S_{y}(0)\cos (\eta t)-\frac{\eta }{\omega _{0}}S_{x}(0)\sin
(\eta t).%
\end{array}%
\right.  \label{S2}
\end{equation}%
where
\begin{equation}
\eta =\sqrt{\omega _{0}(\omega _{0}+q)}.  \label{ET}
\end{equation}%
Based on Eq. (\ref{S2}), we have
\begin{equation}
\left\{
\begin{array}{c}
\Delta S_{x}^{2}(t)=\frac{N}{4}[\cos ^{2}(\eta t)+\frac{\omega _{0}^{2}}{%
\eta ^{2}}\sin ^{2}(\eta t)], \\
\Delta S_{y}^{2}(t)=\frac{N}{4}[\cos ^{2}(\eta t)+\frac{\eta ^{2}}{\omega
_{0}^{2}}\sin ^{2}(\eta t)].%
\end{array}%
\right.  \label{S3}
\end{equation}

Since $\eta >\omega _{0}$, the reduced spin fluctuations occur in the $x$\
direction, i.e., the definition of spin squeezing, $\xi _{\text{R}%
}^{2}(t)=N\Delta S_{\vec{n}_{\bot }}^{2}(t)/\left\vert S(t)\right\vert ^{2}$%
, becomes
\begin{equation}
\xi _{x}^{2}(t)=\frac{4\Delta S_{x}^{2}(t)}{N}.  \label{S4}
\end{equation}%
Substituting the expression $\Delta S_{x}^{2}(t)$ in Eq. (\ref{S3}) into Eq.
(\ref{S4}) and then choosing
\begin{equation}
t=\frac{(2n+1)\pi }{2\omega }\text{ \ \ (}n=0,1,2,\cdots \text{)},
\label{MSFT}
\end{equation}%
the MSF is finally obtained by
\begin{equation}
\xi _{\text{M}}^{2}=\frac{\omega _{0}^{2}}{\eta ^{2}}=\frac{1}{1+q/\omega
_{0}},  \label{MSQ}
\end{equation}%
which agrees with the direct numerical calculation, as shown in Fig. \ref%
{fig6}.

\begin{figure}[tp]
\includegraphics[width=8.5cm]{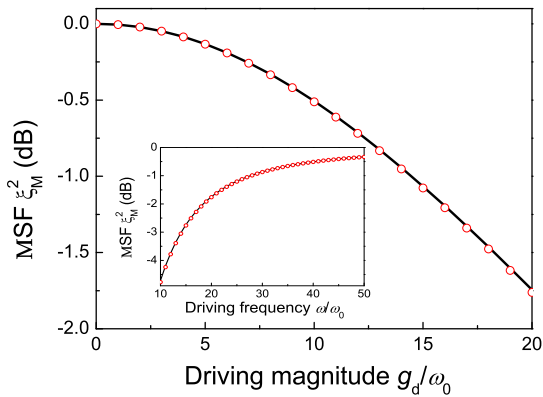}\newline
\caption{The MSF $\protect\xi _{\text{M}}^{2}$ as a function of the driving
magnitude $g_{d}$, when $\protect\omega =20\protect\omega _{0}$ and $N=100$.
Insert: The MSF $\protect\xi _{\text{M}}^{2}$ versus the driving frequency $%
\protect\omega $, when $g_{d}=20\protect\omega _{0}$ and $N=10$. In these
figures, the black solid curves denote the direct numerical calculation of
the effective time-independent Hamiltonian (\protect\ref{HE}), and the red
open symbols correspond to the analytical results in Eq. (\protect\ref{MSQ}%
). }
\label{fig6}
\end{figure}

It can be seen from Eq. (\ref{MSQ}) that, when increasing the driving
magnitude $g_{d}$, the MSF $\xi _{\text{M}}^{2}$ increases (red dashed curve
in Fig. \ref{fig2}), but decreases when increasing the driving frequency $%
\omega $ (Fig. \ref{fig3}). In addition, Eq. (\ref{MSQ}) also shows that the
MSF $\xi _{\text{M}}^{2}$ is independent of the atom number $N$. In fact,
the value of the atom number $N$ restricts the upper limits of the repulsive
spin-spin interaction strength $q$, since Eq. (\ref{MSQ}) is valid for $q\ll
N\omega _{0}$. When $N=10^{4}$, we approximately take $q=10^{3}\omega _{0}$,
which becomes $q=10^{4}\omega _{0}$ when $N=10^{5}$. This means that using
current experimental parameters with $N=10^{5}$, the MSF reaches $40$ dB ($%
30 $ dB for $N=10^{4}$). When $\omega _{0}\ \sim q/N$ or $\omega _{0}\ <q/N$%
, the analytical expression in Eq. (\ref{MSQ}) is invalid. However, with
decreasing $\omega _{0}$, the MSF increases \cite{JM11,NPR13}.

In multicomponent BECs, the spin-spin interactions can also be realized by
controlling the direct atom-atom collision interactions. In principle, this
effective spin-spin interaction can be tuned by a magnetic-field-dependent
Feshbach resonant technique \cite{CC10}. However, similar to the result of
the undriven Dicke model, its strength is also weak (from kHz$\ $to MHz),
and is far smaller than our prediction ($\sim $ several hundred MHz). As a
result, the generated MSF is also far smaller than our result ($40$ dB).

\section{Possible experimental observations}

Here we briefly discuss how to possibly observe these results in
experiments. As an example, we consider the D$_{2}$-line of $^{87}$Rb. The
two stable ground states, $\left\vert G_{1}\right\rangle \ $and $\left\vert
G_{2}\right\rangle $\ in our proposal, are chosen as two hyperfine substates
of $5S_{1/2}$, i.e., $\left\vert F=1,m_{F}=-1\right\rangle =\left\vert
G_{1}\right\rangle $\ and $\left\vert F=2,m_{F}=2\right\rangle =\left\vert
G_{2}\right\rangle $, with a splitting $\sim 2\pi \times 6.8$\ GHz ; whereas
the virtual excited states $\left\vert 1\right\rangle $\ and $\left\vert
2\right\rangle $\ can be chosen as two of the hyperfine substates in the 5P$%
_{1/2}$\ excited state. The decay rates for the 5P$_{1/2}$\ excited state
and the photon are given by $\gamma =2\pi \times 3$\ MHz and $\kappa =2\pi
\times 1.3$\ MHz, respectively \cite{FB09}.

In addition, by controlling the frequency $\omega _{b}^{\prime }$, which is
close to the frequency $\omega _{b}$\ of the energy level $\left\vert
G_{2}\right\rangle $, the effective atom frequency is of the order of
several hundred kHz. Moreover, the repulsive spin-spin interaction strength $%
q$\ can be of the order of a GHz, by independently manipulating the
intensities of the classical driving lasers. For example, when the
parameters are chosen as $g_{d}=\omega =2\pi \times 0.5$\ GHz and $\Delta
_{p}=0.1\omega =2\pi \times 0.05$\ GHz, then $q=\Delta
_{p}g_{d}^{2}/(2\omega ^{2})=2\pi \times 250$\ MHz. Therefore, the condition
$q=10^{4}\omega _{0}$\ for achieving the MSF $\xi _{\text{M}}^{2}=40$\ dB
can be satisfied. Moreover, the shortest time for generating the MSF is $%
t=\pi /(2\omega )=0.5$\ ns $\ll $\ $\tau _{a}=1/\gamma =53$\ ns. This means
that the giant spin squeezing can be well realized within the atom decay
time $\tau _{a}$.

\section{Conclusions}

In summary, we have presented an experimentally-feasible method to achieve a
giant and tunable spin squeezing by introducing a time-dependent collective
atom-photon coupling. We have demonstrated explicitly, in the high-frequency
approximation, that this spin squeezing arises from the strong repulsive
spin-spin interaction induced by the time-dependent collective atom-photon
coupling. More importantly, using current experimental parameters with $%
N=10^{5}$, we have derived a giant MSF of about $40$ dB. We believe that
these results could have applications in quantum information and quantum
metrology.

\section{Acknowledgements}

We thank Prof. Su Yi, and Drs. Qifeng Liang, Jian Ma, and Y.G. Deng for
their useful comments. This work was supported partly by the 973 program
under Grant No. 2012CB921603; the NNSFC under Grant No. 61275211; the PCSIRT
under Grant No. IRT13076; the NCET under Grant No. 13-0882; the FANEDD under
Grant No. 201316; and the ZJNSF under Grant No. LY13A040001. FN is partially
supported by the RIKEN iTHES Project; MURI Center for Dynamic
Magneto-Optics; Grant-in-Aid for Scientific Research (S); MEXT Kakenhi on
Quantum Cybernetics; and the JSPS via its FIRST program.

\end{document}